\documentclass[twocolumn,showpacs,preprintnumbers,amsmath,amssymb]{revtex4}

\usepackage{graphicx}
\usepackage{dcolumn}
\usepackage{bm}
\usepackage{epsfig}

\begin{document}

\title{Reaction-limited Colloidal Aggregation Induced by Salt and Inert Polymers}

\author{ M. Hosek}
\affiliation{
Physics Department \\
Indiana University \\
Bloomington, Indiana \\
}
\author{ J. X. Tang }
\affiliation{
Physics Department \\
Brown University \\
Providence, Rhode Island \\
}
\date{\today}

\begin{abstract}

   Salt-induced aggregation of 
20~nm colloidal silica is followed by light transmission,
which shows an a kinetic form 
$\exp[-(t/t_0)^{\alpha}]$, where
$\alpha = 2.6 $ and $t_0$ is an empirical time constant which
reflects the colloidal stability. 
We found a power law dependence of $t_0$ on ionic strength, 
which can be  explained by  
the classical DLVO theory.
The neutral polymers  polyethylene glycol 
(PEG) accelerate the aggregation rate, and those with higher molecular weight
are more effective in inducing the aggregation
with similar stretched exponential 
form of kinetics.
Current theories of polymer-mediated interactions provide
a reasonable 
interpretation of the effect of  PEG. 
The stretched exponential kinetics of the light transmission is found to be  
consistent with a cluster-size
dynamic scaling model of aggregation.

\end{abstract}


\maketitle

\section{Introduction}

   If a particle in solution is small enough, the gravitational potential energy over a
macroscopic length is less than the thermal energy.  Such colloidal
particles distribute themselves evenly in a vial by thermal diffusion.
If the colloid were to have no surface charge, the particles would
collide,
stick together, and eventually deposit
at the bottom of the vial.  Adequately charged colloidal particles 
repel each other and
stay in suspension essentially forever. 
This property of colloidal stability is of immense practical relevance to 
real world materials such as food, adhesives, cosmetics, inks, and paints.
The silica  nanoparticle is of  
longstanding technological significance \cite{Alexander_book} and
in a sense is the ancestor of the modern nanoparticle.
Today it has uses ranging from  chemical-mechanical polishing of 
silicon wafers to serving as a DNA carrier in non-viral transfection \cite{saltzman-GeneThrpy-v13}.

  Colloidal silica aggregates with the addition of
salt, which lowers the particle surface potential and Debye screening
length.
The presence of a non-adsorbing polymer can also
induce aggregation. This is understood as a "depletion
force", "entropic force", or "molecular crowding" generated by a
contest for space between the colloid and the free polymer.
Asakura and Oosawa (AO) \cite{AandO54,hansen_barrat_book} were the first to interpret
this simple physical concept.
Further theoretical progress has been made over the past decades,
including those using the 
methods of integral equation theory \cite{fuchs-schweizer-JP-v14}
and the scaling/renormalization 
approach \cite{deGennes_book,schafer_book,tuinier-PRE-v65,tuinier-EPJ-v6}.

 Here we study the destabilization  of colloidal silica by 
both monovalent salt and 
the polymer PEG. 
We should mention here that PEG is also of
ascending importance in the  science of biomaterials.
In enzymology, it has been used to study hydration effects, steric hindrance, 
and molecular crowding. When covalently bound to a surface it
serves as a biocompatible passivation layer.

  Charged colloidal particles in saline solution are typically modeled as bodies
with repulsive (stabilizing) Coulomb interactions and attractive short-range
dispersion forces, an approximation known as the classical 
Derjaguin, Landau, Verwey and Overbeek  (DLVO) 
theory \cite{VO_book,israelachvili}.
The electrostatic potential is modeled by the
Poisson-Boltzmann (PB) equation, which describes the relation between charge density
and electrical potential,
under the assumption that the saline ionic
charges deviate from their bulk concentration 
according to Boltzmann's
law.  For moderately charged spherical particles, the external electrostatic
potential is of the form $\phi_0 \exp(-r \kappa)/r$, where $1/\kappa$ is the Debye
screening length.  In a monovalent salt solution, 
$\kappa^2 = {e^2 n}/{ \epsilon \epsilon_0 k_B T}$ where $\epsilon \epsilon_0$ is
the solution dielectric constant and $n$ the number concentration of salt.
In aqueous 1~M monovalent salt solution, for example,  $1/\kappa = 0.3~\rm{nm}$.
Increasing ionic strength also lowers the particle surface potential $\phi_0$.
These two factors combine to decrease the electrostatic interaction potential
$U_{ele}$ and hence lower the stability of a colloidal
suspension.  

  Attractive dispersion forces are the result of induced dipole-dipole interactions;
these are responsible for the irreversible binding in aggregation.
For two volume elements of material the interaction force is 
proportional to $A/r^6$ where $A$ is the Hammaker constant.
The resulting potential is also called the van~der~Waals potential $U_{vdW}$. 
The total interaction is then  $U_{ele} + U_{vdW}$.

   With increased ionic strength the colloidal particles 
undergo irreversible formation of larger and larger clusters,
a process known as aggregation.
It has been shown \cite{brodie-cohn-PRL-v64,lin_PRA-v41}
that the cluster kinetics, size distribution as a function of time,
and cluster fractal dimension fall into two universal classes:
diffusion limited aggregation (DLA) and reaction limited 
aggregation (RLA). In RLA single particles and clusters collide
many times before joining together irreversibly. In the DLA limit, 
particles and clusters aggregate the first time they collide,
so particle diffusion limits the rate of aggregation. 
The clusters generated in these two regimes have different 
qualitative appearance; the fractal dimension of DLA particles
is $d_f \sim 1.7$ and for 
RLA $d_f \sim 2.0$ \cite{brodie-cohn-PRL-v64,lin_PRA-v41,hansen_barrat_book}.
These two classes also exhibit different time evolution scaling laws in 
cluster-size distribution \cite{lin_PRA-v41,brodie-cohn-PRL-v64}. Cluster growth had been
experimentally observed for quite some time \cite{turkevich-JACS-v85}, but did not
have the benefit of a modern conceptual framework.

    As the clusters have fractal dimension less than 3, it is possible for
a small  concentration of particles to span the sample volume.
The cluster radius of gyration $R_g = b n^{1/d_f}$ where
$b$ is the monomer radius and $n$ is the number of
particles in the cluster. 
As the bulk  monomer density 
approaches the cluster  particle density $n/R_g^3$,
the  clusters in effect  span the sample volume.

  There are similarities and differences  between aggregation and
a second-order phase transition. As the process of aggregation
proceeds, the particle size becomes larger and larger.
In a  second-order phase transition, the correlation length (analogous to 
cluster size) diverges, but the correlations are 
dynamic, in that particles are free to leave and rejoin
large clusters, whereas in aggregation the particles
irreversibly join a cluster. 
A salient feature of increasing cluster size is
increasing light scattering.
The scattered light $I(k) \sim n^2 k^{-d_f}$ 
for $k > R_g^{-1} $ \cite{hansen_barrat_book}
where $k$ is the scattering wave vector related to the scattering angle.
The total scattered light ($I(k)$ integrated over $k > 0$)
determines the turbidity of the sample, and it 
has been used to characterize the correlation length at the
critical region in colloid-polymer
phase separation \cite{robert-PRE-v64}.

 The surface of silica in basic aqueous solution is negatively
charged. Silica $\rm{(SiO_2)}$  reacts with 
water to create silanol  $\rm{(SiOH)}$  surface  groups \cite{Iler_book}.
At a high pH,
protons are pulled from the surface, leaving negatively charged 
silane groups $\rm{(SiO^-)}$, thereby lending colloidal stability. 

   A table-top example of the above concepts is
provided by a 4~\%~w/w solution of 20~nm colloidal silica at pH 10.
As the monovalent salt KCl is increased past 200~mM, 
the colloidal particle surface potential and 
electrostatic screening length decrease, and the silica particles begin
to combine into 
larger and larger clusters.
As a result, the turbidity (light scattering) progressively increases.
In the presence of PEG, the
turbidity occurs more readily. 
We elaborate on this simple demonstration to 
further our understanding of colloidal stability and
polymer-mediated interactions. 

   The rest of this paper is as follows.
We first continue with
a more detailed account of the necessary theoretical background.
After a description of the experimental materials and methods, 
the data is presented and interpreted
in terms
of DLVO theory and the two-particle Smoluchowski equation, 
the depletion force,
and the dynamics of cluster-size distribution in the RLA limit.

\section{Theoretical background}

\subsection{The surface charge of colloidal silica}

 Acid titration data of colloidal silica provides some
surprising details about the nature of its surface \cite{bolt-JCP-v61,kobayashi-Lang-v21}.
The results can be qualitatively explained as follows:
At increased pH, the chemical potential of the solution proton 
gas is lower, and the silica surface protons are
able to enter the solution, leaving behind a more negatively charged surface.
With increasing ionic strength,
the electrostatic potential for a given surface charge is
less, so the particle can assume more charge for a given pH.
The relationship between surface charge, surface potential, and
ionic strength is explained with the PB equation.
However, it does not explain the titration data.
By proposing a surface capacitance, 
the Stern model is successful in modeling the
titration data. The details of this calculation are
in Appendix~\ref{app:appA}. In Fig.~\ref{fig:pH-sigma-buffered} 
we present the results of this calculation for 
a solution of 4~\%~w/w 20~nm silica with 140~mM ammonia buffer.
It should be emphasized that this family of curves
accurately models the experimental titration  data~\cite{bolt-JCP-v61,kobayashi-Lang-v21}
for  colloidal silica.
The success of the Stern model
in explaining this  titration data  
lends confidence that we know reasonable values
for the surface charge density.

\hfill
\newline
\begin{figure}[!b]
\includegraphics [scale=0.35,clip=false] {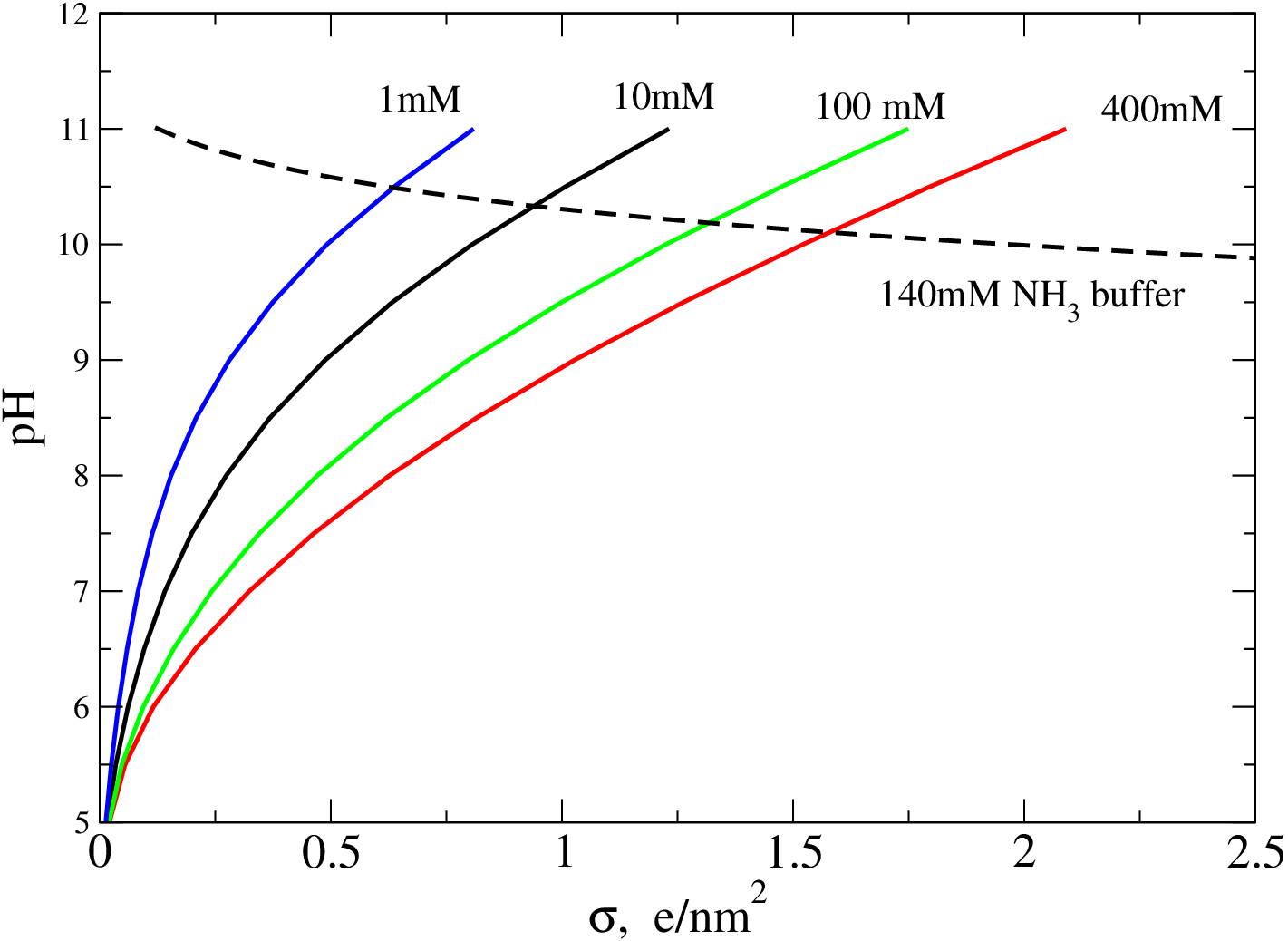}
\newline
\caption{
\label{fig:pH-sigma-buffered}
pH control of $\sigma$ at different monovalent ionic strengths
according to the Stern model of 20~nm silica.
With 4~\%~w/w silica, the buffering strength of 140~mM ammonia ($K_b = 1.77 \times 10^{-5}$M)
is adequate to 
prevent significant change in pH with ionic strength.
Details are explained in Appendix~\ref{app:appA}.
}
\end{figure}
\hfill

\subsection{Two-particle interactions}

 The two-particle interaction is modeled with
a electrostatic repulsion and a short-range 
dispersion attraction, and the polymer-mediated
interaction is treated as an AO depletion attraction.

    We use estimates of the two-particle electrostatic 
interaction $U_{ele}$ 
based on linear superposition 
approximations \cite{carnie-JCIS-v171,RSS_book}: 
\begin{equation}   
           U_{ele}(h) = 32 \pi \epsilon  \epsilon_0 (kT/e)^2 \
      a [\tanh(e \phi_d)]^2 \exp( - h \kappa) \label{U_ele}
\end{equation}
where $a$ is the particle radius, $r$ is the particle center-to-center separation and 
 h  is the surface-to-surface separation $r - 2 a$.
$\phi_d$ is the electrostatic potential at the particle-solution interface.
An essentially identical interaction is obtained
following the work of Behrens and Grier \cite{behrens-grier-JCP-v115}.
In the Appendix it is shown how to obtain the surface charge
density $\sigma$  and hence $\phi_d$ from 
given pH and ionic strength.

  For two spherical particles, the 
van~der~Waals attraction potential is 
$U_{vdW} = -{{A}\over{6}}[ {{2 a^2}\over{r^2 - 4 a^2}} + {{2 a^2}\over{r^2}}
 + \log {{( r^2 + 4 a^2)} \over {r^2}}]$ 
where $A$ is the Hammaker 
constant \cite{VO_book}.  
This formula results from the integration of the 
$A/r^6$ force described in the Introduction.

   For $U_D$ we use the results of the PRISM theory according to
Fuchs and Schweitzer \cite{fuchs-schweizer-JP-v14}
\begin {equation}
U_D =  k_B T {{27} \over {8}}  {{c}\over{c^*}} {{a} \over {R_g}} [ 1 + (5/9)X + (X/3)^2 ] e^{-X}
\end {equation}
where $X = h/\xi_0$.
$\xi_0$ is the polymer mesh
length or equivalently the polymer density-density correlation \cite{fuchs-schweizer-JP-v14}.
In the dilute limit it is the Gaussian radius $\xi_0 = R_g/\sqrt{2}$.
We use  radius of gyration  values $R_g~=~11.4~\text{nm}$ for
35~kD~PEG and $R_g~=~3.1~\text{nm}$ for 4~kD~PEG. 
$c^*$ marks the semidilute polymer concentration where 
the  molecules begin to overlap, \it{i.e.~}\rm $c >  1/(4 \pi R_g^3/3) $.
We use $c^*$ values of  0.92~\%~w/w and  5.25~\%~w/w respectively.

 \subsection {Aggregation and the Smoluchowski equation}

   The aggregation of colloidal particles is a stochastic process. 
For a spherically symmetric configuration the Smoluchowski equation \cite{VO_book,RSS_book} 
provides a continuum description of net flux of two particles:
\begin{equation}
  J = 4 \pi r^2 \big{[} D{{\partial n}\over{\partial r}} + {{n k_B T}\over{2 D}}  {{\partial U(r)}\over{\partial r}} \big{]} 
\end{equation}
where $r$ is the center-to-center particle separation, $n(r)$ is the 
particle density, $D$ is the diffusion constant, and $U(r) = U_{ele} +  U_{vdW} + ... $ is
the particle-particle interaction potential.
This equation is of the form $y' + by = f(x)$ and
can be solved by using the integrating factor $e^{bx}$. 
The boundary conditions 
are $n(2a) = 0$ and $n(\infty) = n_0$. 
That is, the particles fall into a sink
when they touch, and at a large separation the
particle density is the bulk value.
Assuming the quasi-equilibrium condition
of constant $J$, the specific solution is \cite{VO_book}
\begin{equation}
  J = {{4 \pi D n_0} \over {\int_{2a+\delta}^{\infty} \exp\big{[}{U(r)/k_B T}\big{]} d r/r^2 }} \label{J_eqn}
\end{equation}
Here $\delta$ is an arbitrary cutoff value to cope 
with divergence of $U$ at $h=0$.

\subsection { Cluster size  population kinetics }

 Given the high ratio between the observed time scale of turbidity and the
Brownian collision time of 4~\%~w/w silica particles,
$ 3 \eta / (4 k_B T n_0) < 10^{-4}~\rm{sec}$ \cite{VO_book},
the  cluster growth  under consideration is certainly RLA.

\begin{figure}[!h]
\includegraphics [scale=0.35,clip=false] {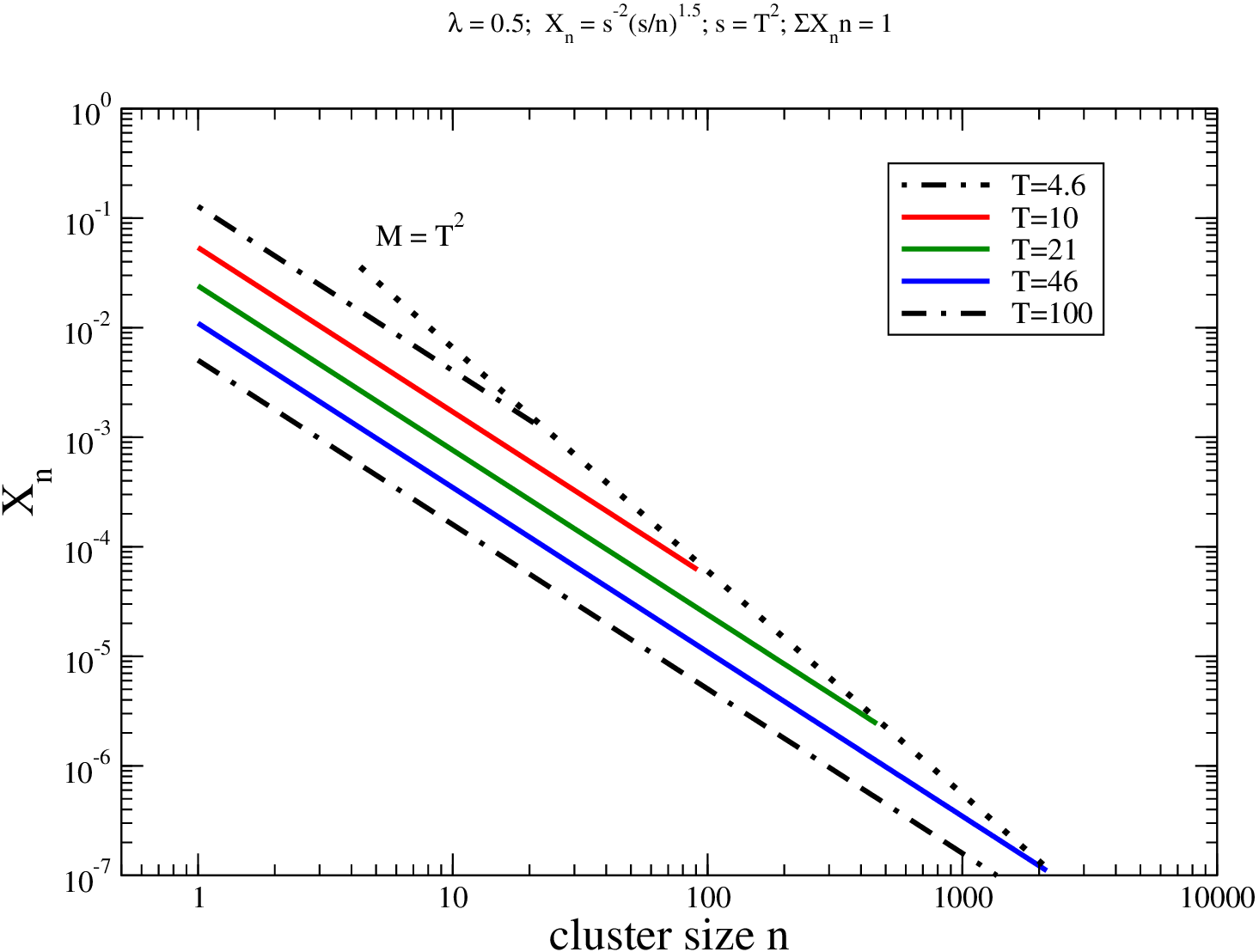} 
\caption{
\label{fig:rla_Xn}
A graphical representation of  Eqns.~(\ref{mass_cons})-(\ref{X_scaling}) 
showing the self-similar nature of the time evolution of
$X_n$. 
In this example $\lambda = 0.5$. $M = T^{2}$ (Eqn.~(\ref{M_time}))  sets the upper limit
of $n$ for each instance of the distribution $X_n(T)$. 
}
\end{figure}

 In the most general scheme, both RLA and DLA  cluster size  population kinetics
can be characterized by dynamic scaling:
$X_n = M^{\theta}f(n/M)$ where $X_n$ is the fraction of
clusters of size $n$, and $M(T)$ is an increasing cluster size 
characteristic of the system at a given time $T$ \cite{OGnFrm_Family,OGnFrm_Leyvraz}, 
where $T = t/t_{agg}$ is time scaled by the
the initial monomer-monomer aggregation rate.
The RLA cluster size dynamics is known to have a power law distribution 
described by the following three equations \cite{OGnFrm_Leyvraz,brodie-cohn-PRL-v64,thorn_PRL_v72}:
\begin{equation}
\sum_{n=1}^{M} n X_n = c_0 = 1   \label{mass_cons}
\end{equation}
\begin{equation}
M(T) = T^{1/(1-\lambda)}             \label{M_time}
\end{equation}
\begin{equation}
X_n = C(T) (M(T)/n)^{1+\lambda}           \label{X_scaling}
\end{equation}
Eqn.~(\ref{mass_cons}) is the constraint of
mass conservation. 
$M$  represents the number of particles in the biggest
cluster of the entire system; it evolves in time
according to Eqn.~(\ref{M_time}).
The third equation shows that $X_n$ 
depends on  $n$ in a power-law fashion with
the characteristic exponent $1+\lambda$.
The scaling exponent $\lambda$
unites the evolution of both $X_n$ and the
the size limit $M$.
$C(T)$ is chosen to satisfy mass conservation.
Given a power law distribution of $X_n$,
the continuum integral $ 1 = \int_1^M n X_n dn$ 
shows that 
$ C(T) = M^{-2}$; it can also be seen that
$M$ is proportional to the cluster size. 
This is summarized graphically in Fig.~\ref{fig:rla_Xn} for
$\lambda = 0.5$.
Eqns.~(\ref{X_scaling}) and  (\ref{M_time}) can be derived from a scaling 
approach to the cluster aggregation process \cite{OGnFrm_Leyvraz}.

\hfill
\newline

 Simulation and experiment in the RLA regime show 
$0.5 < \lambda < 1.0$ \cite{thorn_PRL_v72,gonzales_PRL_v71,family_PRA-v38,rarity-ProcRoySoc-v423,lin_PRA-v41}
and $d_f \sim 2.1$ \cite{cannel-PRL-v52,rarity-ProcRoySoc-v423}.
The asymptotic limit of the scaling theory \cite{thorn_PRL_v72} 
gives $\lambda = 0.5 $.

  Power law growth of $M$ 
is seen experimentally \cite{brodie-cohn-PRL-v64,rarity-ProcRoySoc-v423}  and
by simulations \cite{gonzales_PRL_v71,family_PRA-v38}. 
There is
a delay before the power law growth sets in, but
the resulting $M(T)$ may be time-scaled.

\subsection {Light scattered by a fractal cluster}

   For an individual cluster of size $n$ the 
total scattered light of wavelength $\lambda$ is
\begin{equation}
 n^2 \int S(k,R_g) dk  \label{tau_eqn1}
\end{equation}
where $S(k,R_g)$ is scattering structure factor \cite{ferri-EPL-v7,Fisher_Burford}
and the  scattering vector $k = (4 \pi/\lambda) \sin(\theta/2)$.
For a cluster of fractal dimension $d_f$ 
\cite{ferri-EPL-v7,cannel-PRL-v52,Fisher_Burford,sorensen_PRE}
\begin{equation}
S(k,R_g) = {\Big[1 + {{(R_g k)^2}\over{3 d_f/2}}\Big]^{-d_f/2}}. \label{FB_eqn}
\end{equation}
Thus the total scattering increases
and the low angle scattering becomes greater as 
the cluster size $R_g$ becomes larger. 

     The normalized scattered light $\tau$ from a volume element  
of sample (the turbidity) would then be 
\begin{equation}
 \tau =B \sum_{n=1}^{M} X_n n^2 \int_{k_0}^{k_1} S(k,R_g(n)) dk \label{tau_eqn}
\end{equation}
where $B$ is a constant~\cite{ferri-EPL-v7}.
The spacial derivative of the light intensity $d I /d x \sim -\tau I$, thus
the transmitted light follows a Beer's Law behavior, $ I \sim I_0 \exp(-\tau \Delta x)$ for
arbitrary $\Delta x$.

\section{Materials and Methods}

  In order to follow the aggregation of many samples of 
colloidal silica over a long
time course, a  carrousel was constructed to carry up to 32 standard 3~mL plastic
sample cuvettes with a 1~cm optical path length. 
One cuvette served as the water reference.
An incandescent bulb served at the measuring light
source. Light was collected with a lens and focused into a pencil
which passed through the rotating cuvettes. 
The transmitted light was detected with a photodiode; a
slit in front of this photodetector made a 30~mrad angle
of acceptance.
A picoammeter recorded the photodiode current and was interfaced
with a personal computer.
The computer also controlled the 
rotation of the carrousel, allowing the light transmission to be
recorded for a chosen time schedule.
Temperature
was controlled to be slightly above ambient, 30$^\circ$C.
    20~nm colloidal silica was purchased from Alfa-Aesar
(stock number 12727). PEG (Sigma 03557 and Alfa-Aesar A16151) 
was used as received from the vendors.
Chemicals were of reagent grade.

    To initiate aggregation,  3M KCl was introduced by slow addition (0.1~mL/min)
with a syringe pump to 
the $\sim$~3~mL volume of silica while being throughly mixed with
a magnetically driven stirring propeller.  A glass test tube 
of 1~cm inside diameter
was used to contain the solution while mixing. The solution was then
transfered to a plastic cuvette and capped with tape.
140~mM ammonia was used as the buffer for both the 
silica solution and the 3M KCl stock solution.
Silica concentration was 4~\%~w/w.
By investigating slow aggregation,
practical concerns about the exact initial conditions and
the mixing process were minimized.
The samples were never mechanically disturbed or shaken
after the initial mixing. When PEG was part of the
solution, it was added before the salt. 

    A table-top demonstration of the 
acceleration of aggregation by PEG can be done
by 1:1 v/v mixing of 4~\%~w/w silica with
20~\%~w/w 35~kD PEG.  Both solutions have
200~mM KCl and 140~mM ammonia buffer.
This silica suspension is undergoing aggregation, but very slowly.
By layering a lighter PEG solution on top of
a denser silica solution, a whitish band of 
rapidly aggregating silica is observed at the interface.

    The diffusion constant, which is inversely proportional to
viscosity, plays a role in the Smoluchowski 
model of aggregation.
Viscosity data is available for 2~kD to 6~kD PEG in the
literature \cite{timasheff92}, but the range of concentrations
measured do not extend into the semidilute~\cite{deGennes_book} regime  $c >  c^*$.
Because our measurements extended into this concentration range for 35~kD PEG, whose
semidilute concentration starts at about 0.9~\%~w/w,
the specific viscosities for PEG 4~kD and 35~kD PEG were measured.
Viscosity measurements were
performed with a 30 gauge stainless steel
flow tube at 25$^\circ$C. Water, 23~\%~w/w sucrose, and  46~\%~w/w sucrose
served as calibration standards.
Our results are consistent with the published data  \cite{timasheff92}.
There was no change in the quadratic trend of the specific viscosity
for concentration of 35~kD PEG in the range 0 to 4.5~\%~w/w.
The relative viscosity for 35~kD can be fit with 
$ \eta_{r} = 1.0 +  0.49 c + 0.124 c^2 $ and 4~kD PEG 
with the equation $ \eta_{r} = 1.0 + 0.166 c $ where $c$ is
the \%~w/w concentration in the range 0 to 4.5~\%.

    The radially symmetric PB equation was numerically integrated
as previously described \cite{hosek_tang}. The solutions 
were found to agree with Eqn.~(\ref{sigma_eqn}) for 
relevant values of pH and surface charge. 
The routines \tt{D01GAF~}\rm
and \tt{D01AJF~}\rm from The Numerical Algorithms Group (\tt{http://www.nag.co.uk/}\rm, Oxford UK) 
were used to perform  numerical integrations.

\section{Results}

\subsection{Ionic strength and  aggregation kinetics}

\begin{figure}[!h]
\includegraphics [scale=0.20,clip=false] {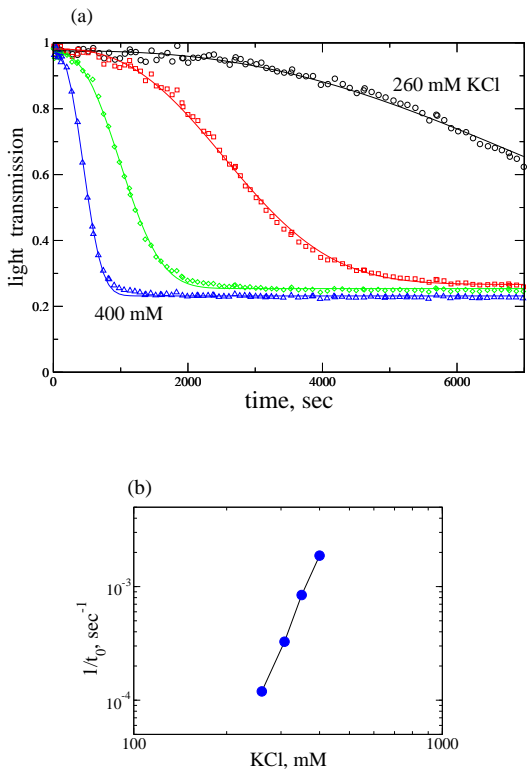}
\caption{
\label{fig:salt_only_data}
(a): Salt-induced aggregation 
followed by change in light transmission for 4~\%~w/w 20~nm silica.
[KCl] = 260, 308, 350, and 400~mM.
Points are the measured light transmission, lines
are a fit of the form $a_0 + a_1 \exp(-(t/t_0)^{2.6})$.
(b): The $1/t_0$ values of the four curves of (a) show a power law
dependence on [KCl] with slope 6.5.  
}
\end{figure}

  Fig.~\ref{fig:salt_only_data} 
presents a data set illustrating the effects of ionic strength on 
the aggregation kinetics of 20~nm silica.
Each time course of the  light transmission 
is empirically described 
with the function
$a_0 + a_1 e^{-(t/t_0)^{2.6}} $.

\hfill
\newline
\begin{figure}[!h]
\includegraphics [scale=0.290,clip=true] {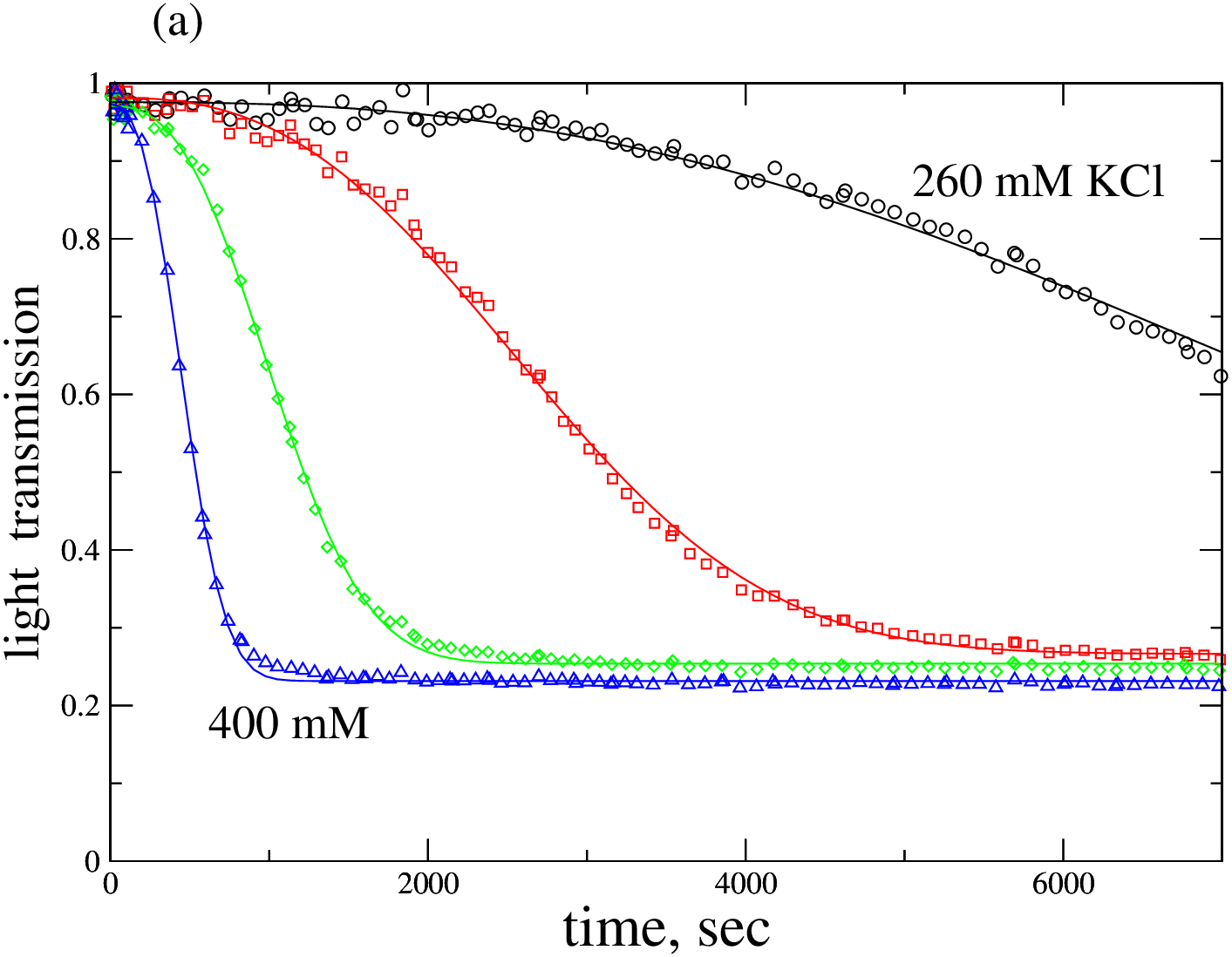}
\end{figure}

\begin{figure}[!h]
\includegraphics [scale=0.290,clip=false] {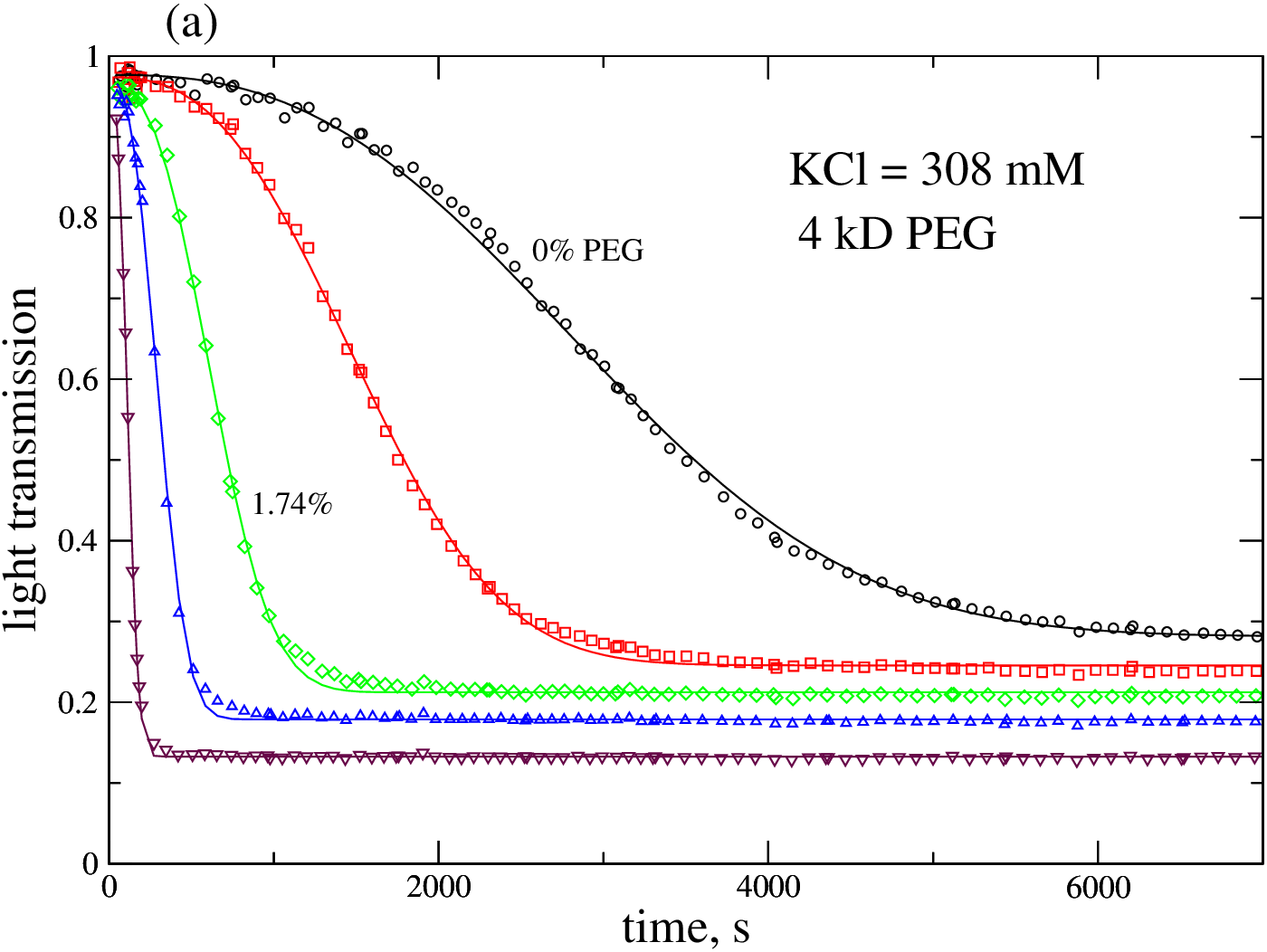}
\end{figure}
\begin{figure}[!h]
\includegraphics [scale=0.290,clip=false] {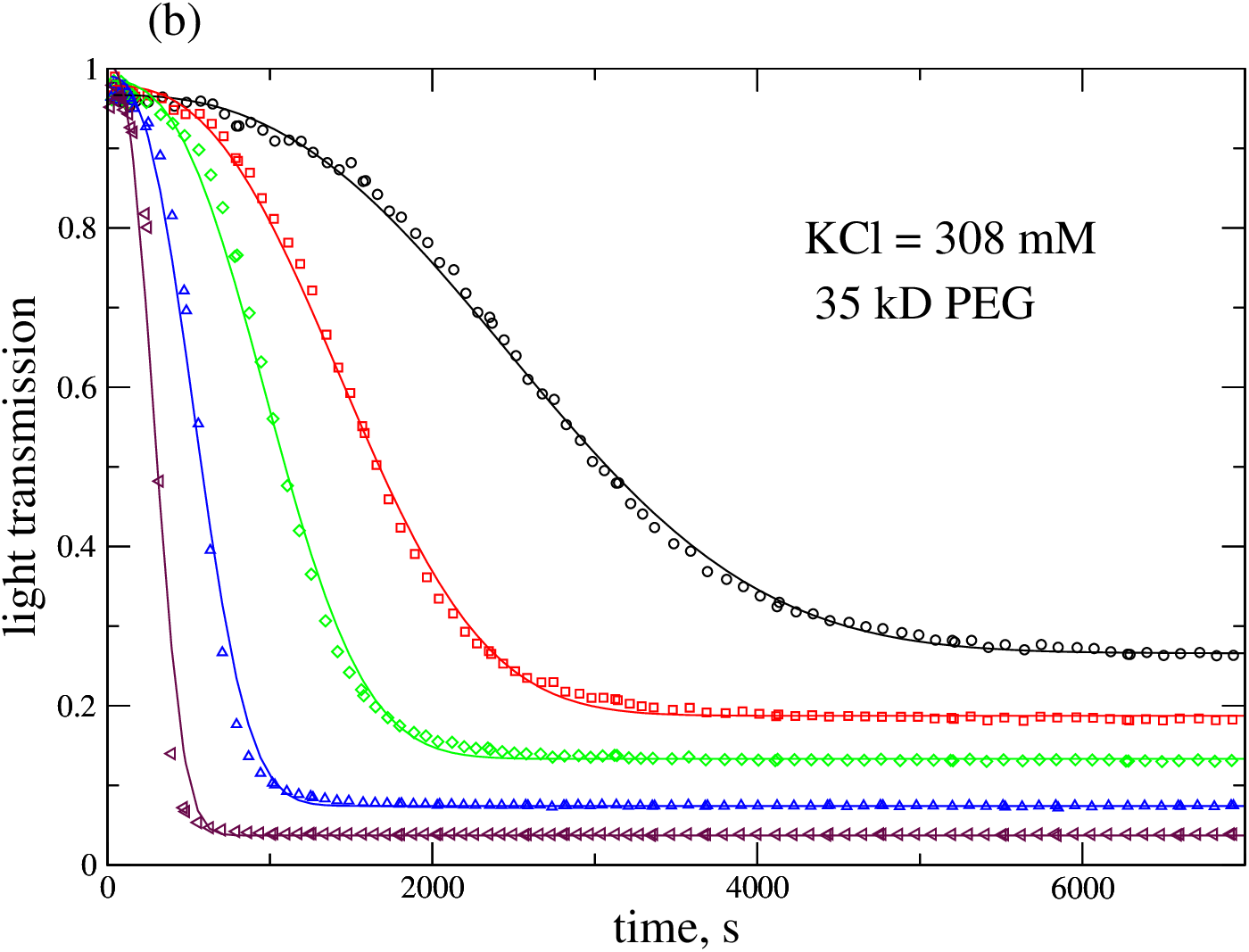}
\caption{ \label{fig:peg_data} The effect of 
4~kD and 35~kD PEG on the aggregation kinetics with 308~mM ionic
strength,
4~\%~w/w 20~nm silica. 
(a): 4~kD PEG [0, 0.87, 1.75, 2.6, and 3.5~\%~w/w];
(b): 35~kD PEG [0, 0.375, 0.747, 1.12, and 1.49~\%~w/w]. 
Points are the data, and lines are the parametric fit
using the same stretched exponential function as in Fig.~\ref{fig:salt_only_data}.
 }
\end{figure}

    We use the salt-induced 
aggregation time courses, shown in Fig.~\ref{fig:salt_only_data}(a), as a means to
evaluate our understanding of the interaction $U(r)$. A log
representation of the dependence of $1/t_0$ on ionic strength (Fig.~\ref{fig:salt_only_data}(b))
demonstrates a power law with an exponent value
of 6.5. Such a power-law relationship has been known for some time to exist
over a limited range of salt concentration and is explained by the
DLVO theory \cite{VO_book}.

 Eqn.~(\ref{J_eqn}) predicts the initial rate of dimer formation 
to be a function of $n_0^2$.  We do observe a quadratic dependence of $1/t_0$ 
on silica concentration (data not shown).

\subsection {Polymer-mediated interaction}

 By setting the aggregation rate at a practical value  with ionic strength, 
we can evaluate  polymer-mediated  effects.
In Fig.~\ref{fig:peg_data}(a) and (b) we demonstrate the
faster development of turbidity caused by  4~kD and 35~kD~PEG, respectively.
Notably, the stretched exponential
power term $\alpha = 2.6$ is preserved for all polymer concentrations.
However, the final turbidity increases with
increasing polymer concentration, whereas the final turbidity 
for the salt-only data is almost constant.
Clearly 35~kD PEG is more effective than 4~kD PEG in accelerating aggregation.

\section { discussion }

\subsection{Kinetics and the two-particle interaction}

\begin{figure}[!h]
\includegraphics [scale=0.30,clip=false] {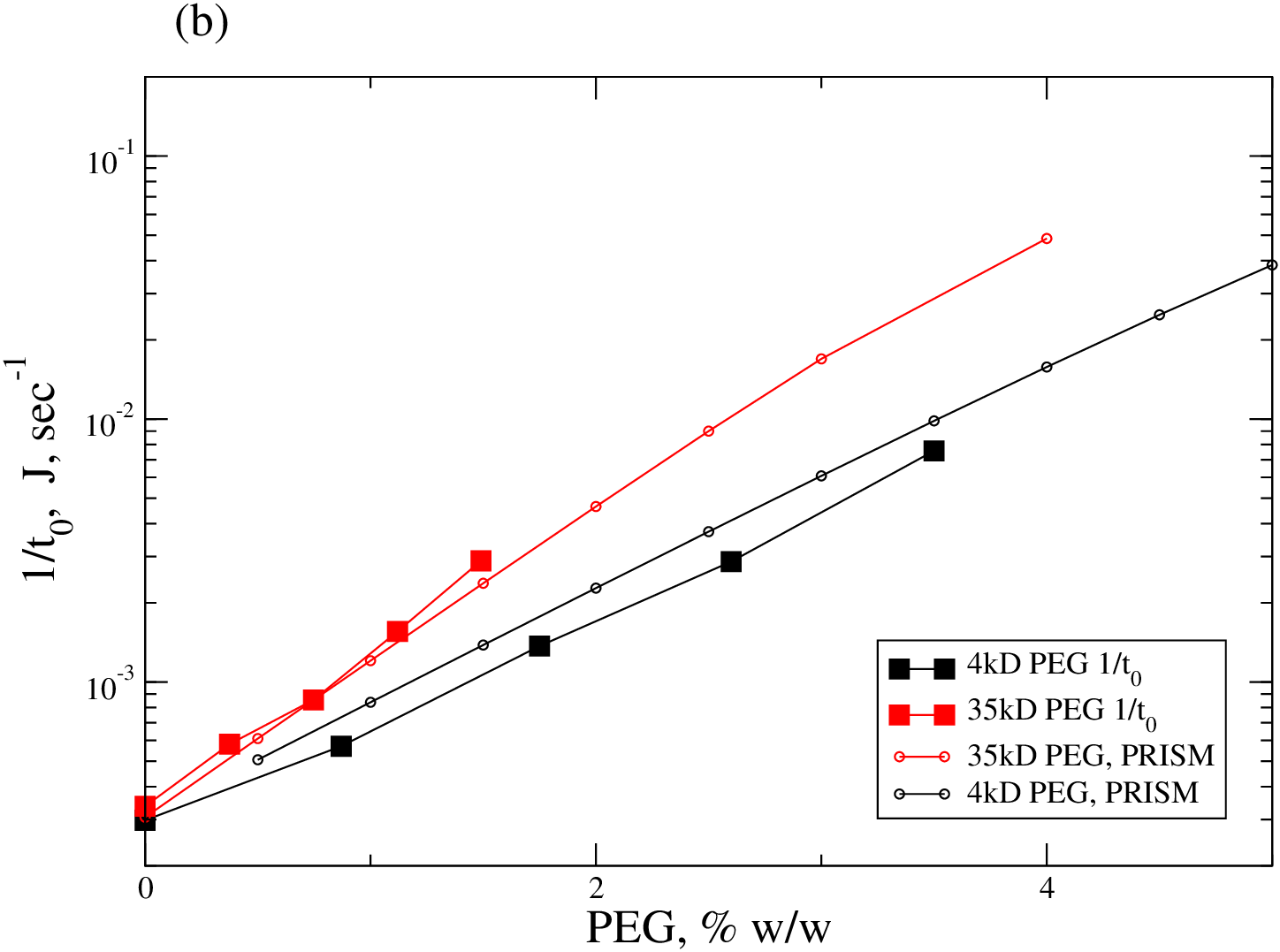}
 \newline
 \caption{
 \label{fig:J-calc-X}
(a): The rate of dimer formation $J$ is extremely
sensitive to the interaction potential. By scaling
$U_1(r)$ by a factor $\gamma = 0.6$, 
$J$ matches the experimental values $1/t_0$ from
Fig.~\ref{fig:salt_only_data}.
(b): Calculation of $J_{eff}$ using PRISM results.
Data from Fig.~\ref{fig:peg_data} is included.
The results of the calculation were scaled 
to match the data at 
zero polymer concentration.  The PRISM interaction  is 
from Ref.~\cite{fuchs-schweizer-JP-v14}, equations (8) and (12).
}
 \end{figure}

As the
integrand in Eqn.~(\ref{J_eqn}) is of the form 
$\exp(U(r)/k_BT)$,
it is expected that the electrostatic interaction will have a
strong effect on the aggregation rate.
The dramatic effect of scaling the interaction by a constant 
is seen in Fig.~\ref{fig:J-calc-X}(a).
It is found empirically 
that an adjusted interaction 
\begin{equation}
U_1 = 0.6\big[U_{ele} +  U_{vdW}\big]  \label{U_eqn}
\end{equation}
demonstrates  a reasonable agreement 
with the $1/t_0$ values of Fig.~\ref{fig:salt_only_data}(b).
These numerical integrations
of Eqn.~(\ref{J_eqn}) use the interaction potential $U_{ele} + U_{vdW}$, with 
$U_{ele}$ based
on the values for $\sigma$  derived from Fig.~\ref{fig:pH-sigma-buffered}.
A cutoff value $c = 0.2$~nm was used in Eqn.~(\ref{J_eqn})
as $U_{vdW}$ diverges at $r=2a$.
We use a Hammaker constant  
$A = 1.66~k_B T$, which is close to the 
commonly accepted Hammaker constant for silica particles in water
of $ 2~k_B T$ \cite{israelachvili}.

 The depletion interactions
predicted by the PRISM integral equation theory
were added to $U_1$, and the resulting $J$ compared with 
the experimental values of $1/t_0$ from Fig.~\ref{fig:peg_data}. 
Calculations were also performed using the
the RG theory according to Tuinier \it{et~al.}\rm\cite{tuinier-PRE-v65}.
Similar results were obtained (not shown).

  Both the RG and PRISM theories seem to overestimate the
interaction.
The experimental results may be matched by using  
0.6 as a constant scaling factor. In summary
\begin {equation}
U(r) = 0.6 \big[ U_{vdW} + U_{ele} + U_D \big]
\end{equation}
\begin {equation}
 J_{eff} = J[U(r)] / {\eta_r}  \label{J_eff}
\end{equation}
where $J[U(r)]$ is numerically obtained from Eqn.~(\ref{J_eqn}).
The effective rate of aggregation $J_{eff}$ includes a factor of 
the relative viscosity, which is
a function of the PEG concentration. 
In Fig.~\ref{fig:peg_data}(b) there is 
a rough agreement  between the calculated aggregation rate 
and the measurement.  

 Because there is no indication of Kramer inversion \cite{hansen_barrat_book}
we believe that using macroscopic relative viscosity 
on the microscopic scale is proper.

\begin{figure}[!h]
\includegraphics [scale=0.30,clip=false] {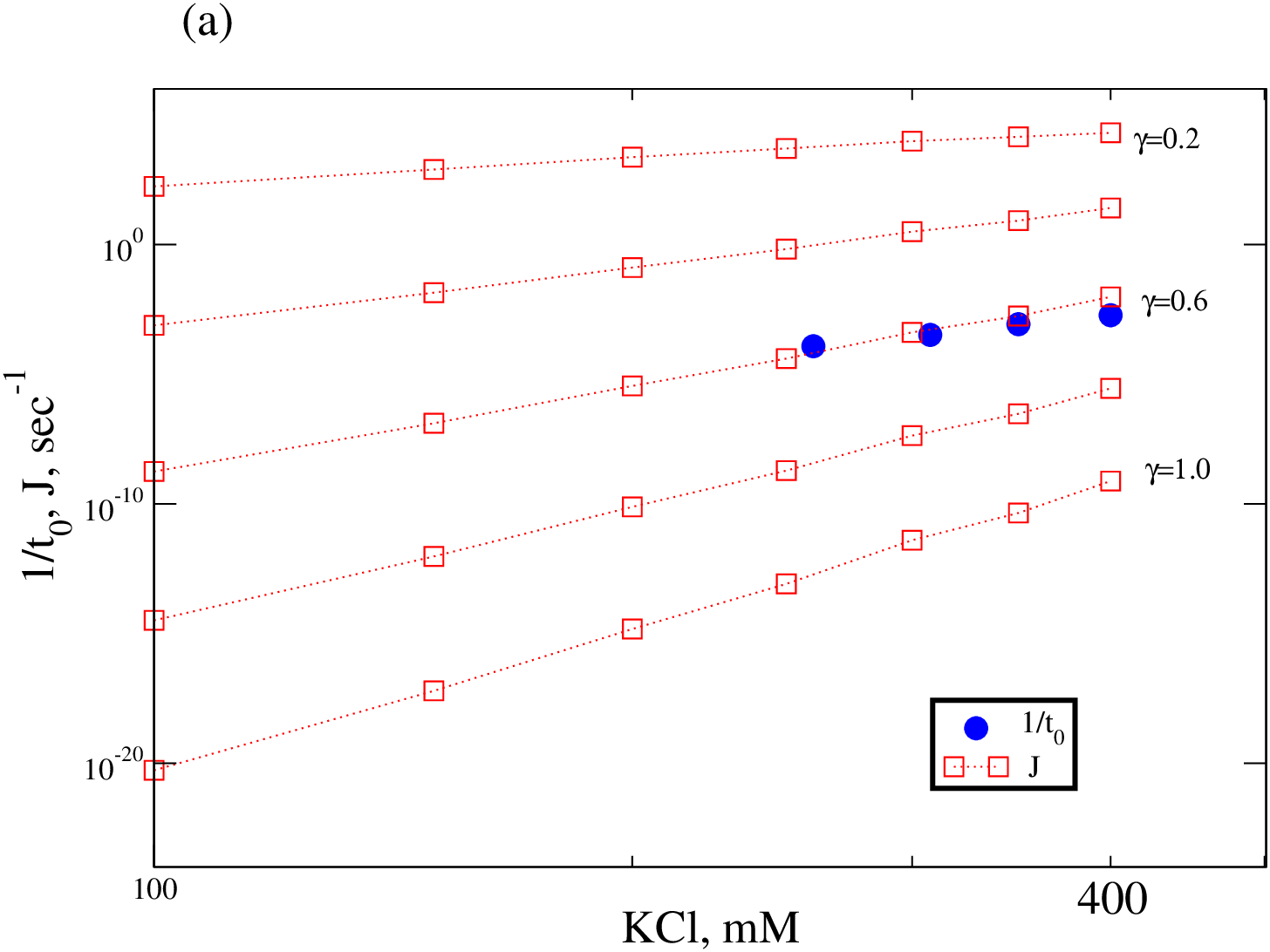}
\end{figure}
 \hfill

\subsection{Origin of the stretched exponential}

\begin{figure}[!h]
\includegraphics [scale=0.30,clip=false] {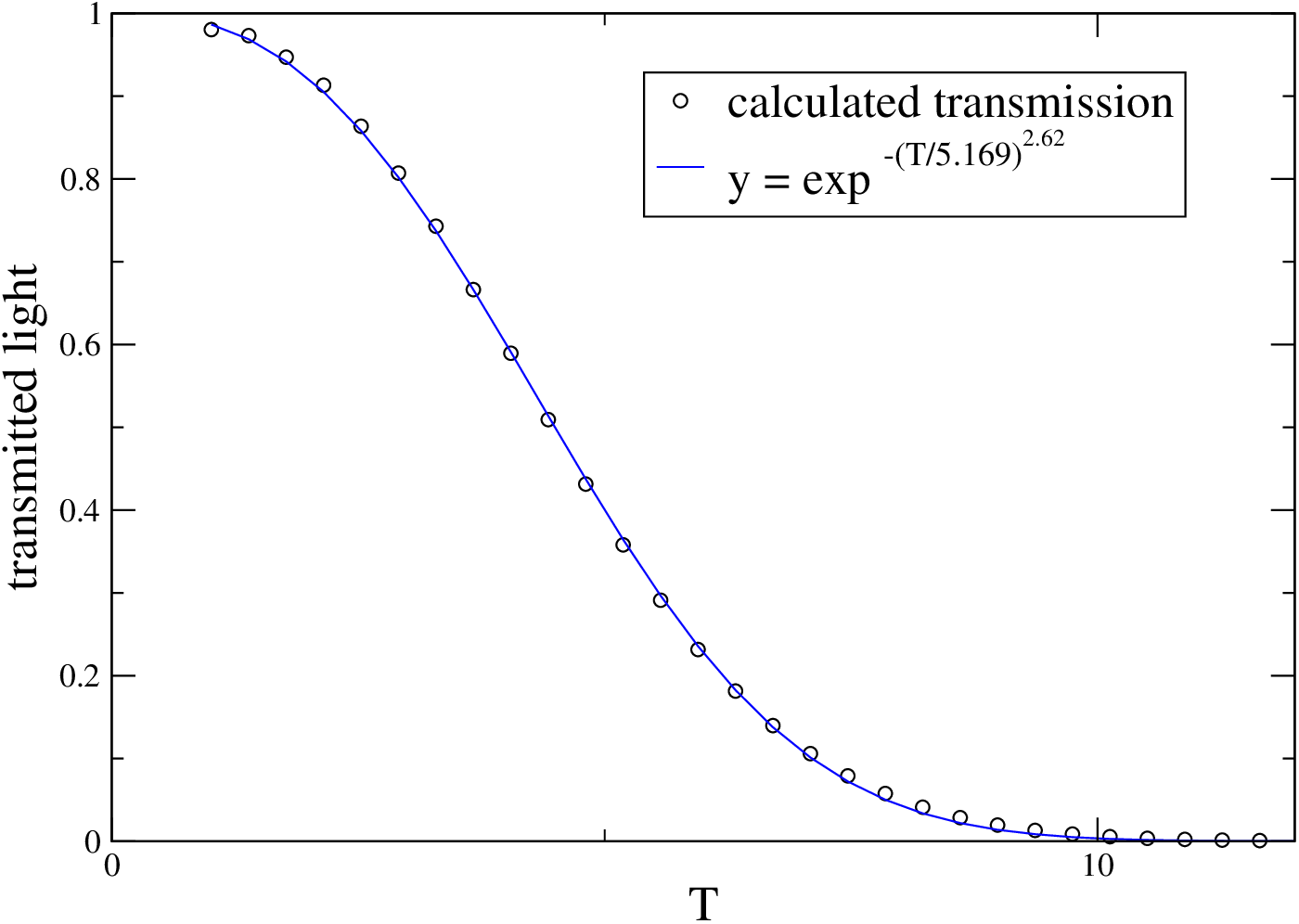}
\caption{
\label{fig:rla_tl}
Light transmission calculation with $\lambda = 0.70$ and $d_f = 2.2$
is able to reproduce  the stretched exponential form of the data ($\exp(-(T/T_0)^{\alpha})$).
In this case $T_0 = 5.16$ and $\alpha=2.6$. The kinetics for other values 
of $\lambda$ and $d_f$ are
listed in Table~\ref{tab:table1}.
Light transmission is proportional to $\exp(-\tau)$.
}
\end{figure}

 The  stretched exponential decay of light transmission
can be explained by combining the analysis of parts C and D.
Numerical evaluation of $\tau$, Eqn.~(\ref{tau_eqn}),
with limits of integration $k_0 = 0$,
$k_1 = 2 \pi/4$,  $d_f = 2.2$, and $\lambda = 0.70$, 
with the cluster size distribution 
of Eqns.~(\ref{mass_cons}),~(\ref{X_scaling}),~and~(\ref{M_time})  
yields the light transmission $\exp(-\tau)$,
This calculation, shown in Fig.~\ref{fig:rla_tl},
closely resembles a stretched exponential form with $\alpha = 2.6$
quite well.
Various stretched exponentials may  be obtained 
from different values of $\lambda$ and $d_f$ (Table~\ref{tab:table1}).  We find the 
kinetics of light transmission is  set primarily by $\lambda$, not $d_f$. 
This can be understood by evaluation of the term $\int S(k,R_g)~dk$ in Eqn.(\ref{tau_eqn}).
This integral is much more sensitive to $R_g$ than $d_f$, and the distribution of
$R_g$ is set by $\lambda$.
\hfill
\newline
\hfill

\begin{table}[!h]
\caption{\label{tab:table1}effect of $d_f$ and  $\lambda$ 
on fit parameters $T_0$ and $\alpha$ for the 
light transmission fitting function
$\exp^{-(T/T_0)^\alpha}$. }
    \begin{tabular}{|c|c|c|c|}
    \hline
   $    $ & $\lambda=0.65$ & $\lambda=0.70$ & $\lambda=0.75$ \\
    \hline  
   $d_f= 2.1$ & $\alpha=2.21$ &  $\alpha=2.55$ &  $\alpha=3.01$ \\
   $    $  & $T_0 =6.72  $ & $T_0 =5.25 $ & $T_0 =4.07 $ \\
    \hline
   $d_f= 2.2$ & - &  $\alpha=2.62$ & - \\
   $    $   & - & $T_0 =5.16 $ & - \\
    \hline
   $d_f= 2.3$ &  $\alpha=2.33$ &  $\alpha=2.68$ &  $\alpha=3.18$ \\
   $    $       & $T_0=6.5$ & $T_0 =5.1 $ & $T_0 =3.9 $ \\
    \hline
    \end{tabular}
\end{table}

 The calculations indicate  zero light transmission as
the aggregation proceeds, whereas 
the data, especially the salt-only data, show a 
limit in turbidity even after gelation is reached.
This is certainly diffusion of
photons in the medium \cite{durian-PRE-v50} which we do not
take into account.

 We have used three time scales here. $t_0$ is the
empirical fitting parameter for the measured light transmission; 
$T$ is the dimensionless time of the
dynamic cluster growth model;  and then $1/J$ is the
dimer formation time scale from the DLVO/Smoluchowski theory.
Simulation \cite{thorn_PRL_v72}
as well as numerical evaluation of the scaling model (Eqns.~(\ref{mass_cons})-(\ref{M_time}))
indicates $X_2 \sim 1/T$, \it{i.e.}\rm, the time scale of
dimer formation is basically $T$. 
So we can reasonably conclude $T/5 = t/ 5 t_{agg} \sim t/t_0 $.
The extreme sensitivity of $J$ 
to the two-particle interaction potential  (as seen in Fig.~\ref{fig:J-calc-X})
implies a rough accuracy in time scale is adequate to characterize
the physics,
so we can 
treat $t_0$ and $t_{agg}$ as the same. 

\subsection {Polymer effects and $\lambda$}

 For salt-induced aggregation (Fig.~\ref{fig:salt_only_data}) it 
appears the process is primarily characterized by $t_0$. Presumably
the geometry of the aggregation (the cluster distribution and
fractal dimension) at a given turbidity is the same for all ionic
strengths; the ionic strength is simply setting the 
rate at which the process is played out.
The nearly constant final turbidity implies the sample
is proceeding to one common state regardless of salt
concentration.

  The effect of PEG is more complex.
Whereas salt-induced aggregation has
only a small difference in turbidity at long time (Fig.~\ref{fig:salt_only_data}), 
PEG-induced aggregation
shows increased final turbidity with 
increased PEG concentration (Fig.~\ref{fig:peg_data}).
But remarkably the interplay of
$\lambda$ and $d_f$ seems to preserve the time-scaling of the turbidity.
According to Ball \it{et al.}\rm \cite{ball-PRL-v58},
$\lambda$ is stabilized by the adjustment of $d_f$.
To paraphrase:
Imagine an increase in $\lambda$, which leads to 
relatively more small clusters, as stated in Eqn.~(\ref{X_scaling}). 
These smaller clusters interpenetrate 
the larger clusters, which combine to form more compact 
objects of higher $d_f$.
The resulting decrease in available surface area 
slows the growth, \it{i.e.}\rm, decreases $\lambda$, which is
consistent with Eqn.~(\ref{M_time}).

\subsection {Flavors of RLA}

  Experimentally $\lambda = 0.5$ has been found with 
very small colloid volume fractions, but with
low colloidal stability set by either divalent ions
or high concentration of NaCl \cite{lin_PRA-v41,brodie-cohn-PRL-v64}. 
In these experiments,
clusters seldom meet, but combine when they do meet.
In our case, and in other experiments \cite{rarity-ProcRoySoc-v423, sorensen_PRE},
NaCl is about 100~mM, so clusters
can meet many times before combining, 
and we find $\lambda = 0.70$.
Two flavors of RLA are consistent with the analysis of 
Meakin and Family \cite{family_PRA-v38}.

 It should be noted that 20~nm silica particles have
a "hairy" nature presumably due to dangling
poly(silicilic acid)  chains on the surface \cite{kobayashi-Lang-v21}.
Larger silica particles are seen to be "harder" and aggregates formed
from such particles show a well-defined particle 
morphology \cite{kobayashi-Lang-v21}.
One might expect these smaller silica particles to form 
"large floppy clusters" \cite{family_PRA-v38} which have
$\lambda = 0.70$.

\section { conclusions }
 To our knowledge this is the first study to combine the
classical DLVO interpretation of colloidal stability with
cluster-size dynamic scaling to further understanding of
the AO depletion interaction. 

 We are able to explain both the stretched exponential 
nature of the light transmission data and the AO depletion interaction
by qualitative
adaption of current theories.

\section {Acknowlegements}
We wish to thank Prof.~John Carini for helpful
comments.

\appendix

\section{The Stern Model} \label{app:appA}  

 Here we explain in detail a model which explains the
titration data of colloidal silica as typified by Fig.~\ref{fig:pH-sigma-buffered}.
It combines the PB theory of electrolyte
solutions, the Stern theory of surface 
capacitance, and the lattice model of the silica surface.

  In a solution of  monovalent salt of concentration $n_0$, 
the electrostatic potential surrounding 
a body with surface charge $\sigma$ can be found by integrating the 
PB equation:
\begin{equation}
\nabla^2 \phi ={2 n_0 e  \sinh ( {{\phi e} / {k_B T} })} / {\epsilon}. \nonumber
\end{equation}
with the boundary conditions  $\sigma / \epsilon \epsilon_0 = \partial \phi(a) / \partial r $
and $\phi(\infty) = 0 $.
By solving this equation, we can find the
surface charge as a function of surface potential $\phi_d = \phi(r=a)$,
particle radius $a$, and Debye screening 
length $\kappa^{-1} \sim \sqrt{n_0}$ \cite{behrens-grier-JCP-v115}:
\begin{align}\label{sigma_eqn}
\sigma(\phi_d) = {{2 \epsilon \epsilon_0 \kappa k_B T } \over {e} } \big{[}  \sinh( e \phi_d / 2 k_B T )  \\ 
\notag \quad \quad + { {2} \over {\kappa a}} \tanh(  e \phi_d / 4 k_B T ) \big{]} 
\end{align}
Because $\sigma$ monotonically increases with $\phi_d$, we
can also readily obtain $\phi_d(\sigma)$.

\hfill
\newline
\begin{figure}[htp]
\includegraphics [scale=0.35,clip=false] {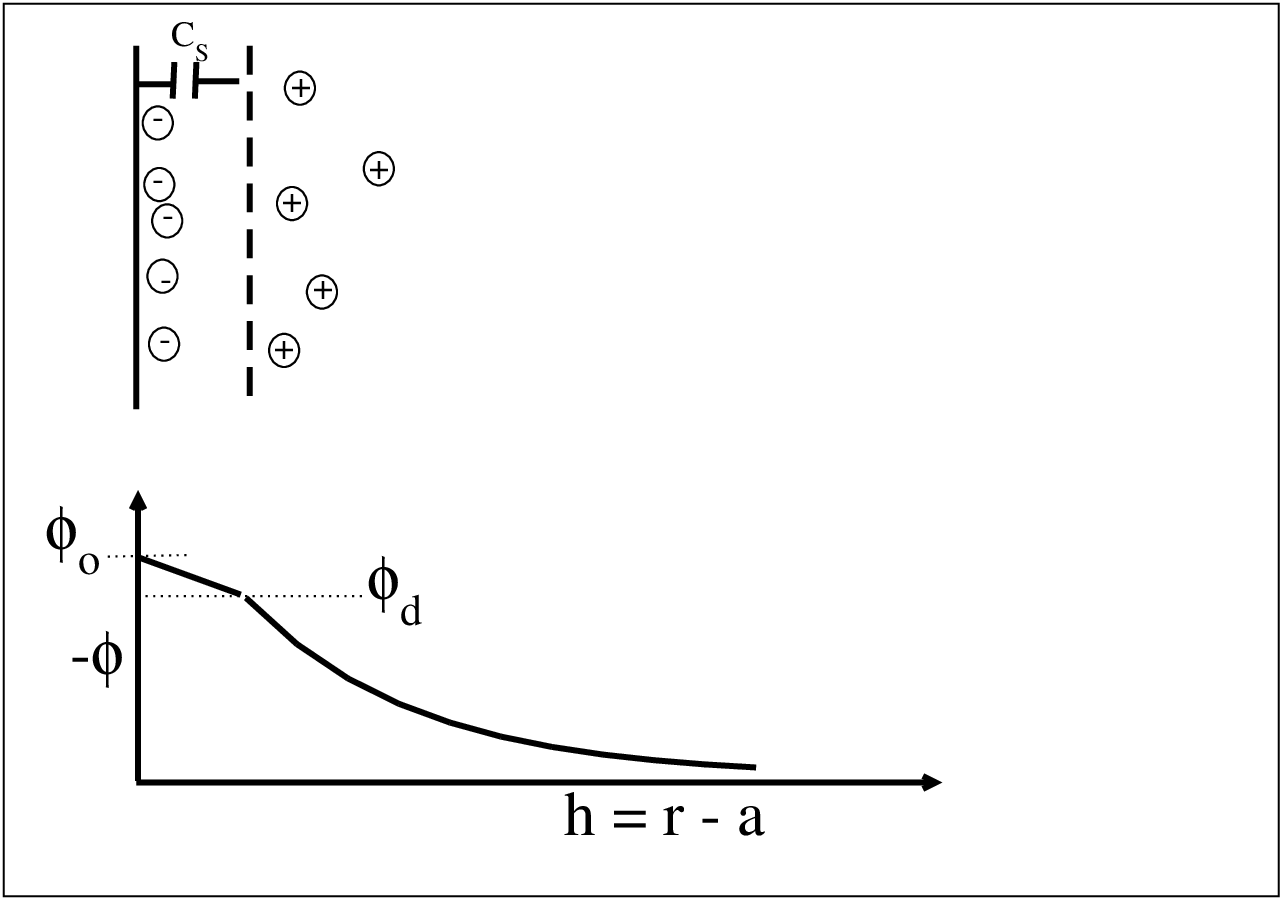}
\newline
\caption{
\label{fig:stern-fig}
The Stern scheme of surface electrostatics.
The (negative) surface charge $\sigma$ experiences potential $\phi_0$. There is
a potential drop $\sigma/C_S$ in going from the surface proper to
the "diffuse" region where the counter-charge dwells. The 
potential experienced in the electrolyte 
medium is described strictly by $\phi(r=a) = \phi_d$ and surface charge $\sigma$.
The Stern layer is merely characterized by
the capacitance/area $C_S$. It can be of arbitrary thickness.
}
\end{figure}

  Without specifying an 
exact physical mechanism, the Stern model proposes
a capacitance $C_S$ between the inner charged layer
and the solution phase.
The potential $\phi_0$ of the surface charge is then 
\begin{equation}
\phi_0 = \phi_d  + \sigma/C_S.  \label{stern_eqn}
\end{equation}
$\phi_d$ is the apparent surface potential 
experienced by the surrounding solution; $\phi_0$ 
is the potential of the charged $\rm{SiO^-}$ surface groups.

     We find it instructive to 
write the  free energy $F$ for one silica sphere in
equilibrium with an ideal gas of protons in solution \cite{carlsson_03}:

\begin{align}
F & = E_{ele} + E_{bind} - TS  \\
\notag & = \int_{q=0}^{q=e(N-n)} \phi_0(q) dq +   n u_0    \\ 
\notag & ~~~~ - T k_B \log {{N!}\over{(N-n)!n!}}  \\ 
\notag & ~~~~ + k_B T (M-n) \log (((M-n)/V)/H_0).   
\end{align}
where there are $(N-n)$ $\rm{SiO^-}$ sites on the silica surface; $n$ is the
number of bound protons forming silanol (SiOH) sites.
The successive terms are: the electrostatic charging energy;
the binding energy for $n$ protons, each forming a silanol group;
the configuration entropy for $n$ indistinguishable 
protons distributed on $N$ possible surface sites;
and lastly,  $M - n$ protons in solution are
treated as an ideal gas of volume $V$.  
This is simply
Langmuir adsorbtion with an electrostatic term. 
$F$ is clearly a function of~$n$, and by minimizing $F(n)$ we
find the surface charge $(N-n)e$.

   This Stern model, with free parameters
density of surface sites $N/4\pi a^2$, proton binding energy $u_0$,
and 
surface capacitance $C_S$,
is quite effective in explaining
the pH titration data of colloidal silica particles for varying monovalent ionic 
strength \cite{bolt-JCP-v61,kobayashi-Lang-v21}.
The results shown in Fig.~\ref{fig:pH-sigma-buffered} represent the titration
curves for several monovalent ionic strengths obtained by
numerically minimizing $F$ using
the accepted values of
$\rm{8~silane~sites/nm^2}$, $ u_0 = $~pK$~k_B T = 7.5~k_B T$,
and $C_S = \rm{2.9~F/m^2}$~\cite{bolt-JCP-v61,kobayashi-Lang-v21}.


  The energetics of the silica charged surface is typically
dealt with in terms of chemical potential \cite{behrens-grier-JCP-v115}.
At equilibrium, the chemical potentials 
of a proton in the gas (solution) phase and 
a proton on the silica  surface  are identical, or equivalently 
$ {{\partial F} / {\partial n}} = 0 $.
Also, the  practical terms $ {\rm{pH}} = - \log_{10} [{\rm{H^{+}}}]$ and
${\rm{pK}} = u_0/k_B T $ are normally employed, hence the term
"one-pK Stern model."

  In summary, for the given pH and 
ionic strength $n_0$, 
we can use the Stern model to find $\phi_d$ and in turn
calculate the particle-particle electrostatic interaction
$U_{ele}$ of Eqn.~(\ref{U_ele}).
\newline

\bibliography{peg-silica-paper}
\end{document}